\documentclass{iau-JDSS}
\usepackage{graphicx}

\title[PMS and MS objects] 
{Early-type PMS and MS objects in M16 and the Carina
star-forming regions}

\author[Martayan et al.]   
{C. Martayan$^{1,2}$, M. Floquet$^2$, Y. Fr\'emat$^3$, A.-M. Hubert$^2$, C. Neiner$^2$, D. Baade$^4$, \and J. Fabregat$^5$
 }
\affiliation{$^1$ ESO, Alonso de Cordova 3107, Vitacura, Casilla 19001, Santiago 19, Chile \break email: cmartaya@eso.org \\[\affilskip]
$^2$ GEPI, Observatoire de Paris, CNRS, 5 place Jules Janssen, 92195 Meudon Cedex, France\\
$^3$ Royal Observatory of Belgium, 3 avenue circulaire, 1180 Brussels, Belgium\\
$^4$ ESO, Karl-Schwarzschild-Str.\ 2, 85748 Garching b.\ M\"unchen, Germany \\
$^5$ Observatorio Astron\'omico de Valencia, Poligon la Coma, 46980 Paterna Valencia, Spain}

\pubyear{2009}
\volume{Volume 15}  
\pagerange{119--126}
\date{?? and in revised form ??}
\jname{Highlights of Astronomy, Volume 14}
\editors{Ian F Corbett, ed.}
\begin{document}

\maketitle

\begin{abstract}
Thanks to a variety of pertinent wide-angle facilities (WFI-slitless mode,
VLT-FLAMES (\cite[Pasquini et al. 2002]{Pasquini et al. 2002}), SPITZER, 2MASS) 
it is possible to comprehensively study the nature
of early-type objects in star-forming regions like the Eagle Nebula and Carina
on large spatial scales.  In them, the young open clusters NGC 6611, Trumpler
14, Trumpler 15, Trumpler 16, and their vicinities are of particular interest. 
With the WFI in its slitless mode (\cite[Baade et al. 1999]{Baade et al. 1999}), 
one can reliably and with little extra effort
discriminate in thousands of spectra between intrinsic circumstellar emission as
in HBe/Ae stars and diffuse interstellar line emission.  The only bias results
from the need of the equivalent width and absolute strength of the line emission
to be sufficient for detection.  VLT-FLAMES spectra combined with infrared data
from SPITZER and 2MASS permit the nature of the objects with and without
emission-lines to be derived. Following this approach, we report on the
discovery and classification of new Herbig Be/Ae stars, pre-main sequence
objects, and main sequence stars in these regions.  Based on line-width
measurements in VLT-FLAMES spectra, the evolution of the rotational velocities
between pre-main sequence and main sequence phases is also discussed. 
\end{abstract}

\firstsection 
\section{Summary of the results}

 We found 11 emission-line stars in NGC6611-M16 and its vicinity, 9 of them are new
Herbig Ae/Be stars.   In the open clusters Trumpler 14, 15, 16, Collinder 232
and their vicinity in the Carina nebula; 6 of the 16 emission-line stars seem to
be  Herbig Ae/Be stars. For  the 10 of the remaining emission-line stars, the
status is uncertain for 2 of them and 8 of them are actually main sequence stars
so are probably  classical Be stars. 

We also found, in agreement with the theory (\cite[Maeder \& Meynet 2001]{MaederMeynet 2001}), that the rotational velocity
decreases by 20\% at the ZAMS between the pre Main Sequence and the Main Sequence
phases. 
All the details about NGC6611 are available in \cite[Martayan et al. (2008)]{Martayan et al. (2008)}.

\end{document}